\documentclass[fleqn]{article}
\usepackage{graphicx}
\usepackage{epsfig}
\newcommand{\AmS}{{\protect\the\textfont2
  A\kern-.1667em\lower.5ex\hbox{M}\kern-.125emS}}

\title{Low Energy Neutrino Physics\\After SNO and KamLAND}

\author{L.\ Oberauer
\smallskip
Technische Universit\"at M\"unchen, Physik E15,\\ 
James-Franck-Str.\ 1, 85748 Garching, Germany}

\begin{document}
\maketitle

\begin{abstract}
In the recent years important discoveries in the field of low energy neutrino physics (E$_\nu$ in the $\approx$ MeV range) have been achieved.
Results of the solar neutrino experiment SNO show clearly flavor transitions from $\nu_e$ to $\nu_{\mu,\tau}$.
In addition, the long standing solar neutrino problem is basically solved.
With KamLAND, an experiment measuring neutrinos emitted from nuclear reactors at large distances, evidence for neutrino oscillations has been found.
The values for the oscillation parameters, amplitude and phase, have been restricted. 
In this paper the potential of future projects in low energy neutrino physics is discussed.
This encompasses future solar and reactor experiments as well as the direct search for neutrino masses.
Finally the potential of a large liquid scintillator detector in an underground laboratory for supernova neutrino detection, solar neutrino detection, and the search for proton decay $p \to K^+ \nu $ is discussed.
\vspace{1pc}
\end{abstract}


\section{Situation after SNO and KamLAND}

In SNO (Sudbury Neutrino Observatory) the flavor transition of solar neutrinos from $\nu_e$ into the $\nu_\mu$ or into the $\nu_\tau$ state has been clearly demonstrated\cite{SNO}.
About 2/3 of the solar $^8$B-neutrinos above $\approx 5$ MeV energy transform into the other flavor on their way from the solar center to the Earth.
The total measured flux of solar $^8$B-neutrinos however is in agreement with theoretical expectations.
Hence, the long standing solar neutrino problem is basically solved.
We know, that the pp-cycle is the main energy source of the sun and the experimental uncertainty on the $^8$B-flux is already below the theoretical one.

In KamLAND the disappearance of $\bar{\nu_e}$, emitted by nuclear power reactors at distances of several hundreds of kilometers has been observed\cite{KamLAND}.
This can be interpreted as evidence for neutrino oscillations, which are a direct consequence of neutrino masses and neutrino mixing.
The weak interaction eigenstates $\nu_\alpha$ ($\alpha = e, \mu, \tau$) can be expressed as linear combinations of mass eigenstates $\nu_i$ ($i=1, 2, 3$)

$$ \pmatrix{\nu_e \cr \nu_\mu \cr \nu_\tau } = \pmatrix{ U_{e1} & U_{e2} & U_{e3} \cr U_{\mu 1} & U_{\mu 2} & U_{\mu 3} \cr
U_{\tau1} & U_{\tau2} & U_{\tau3} \cr} \times \pmatrix{\nu_1 \cr \nu_2 \cr \nu_3 \cr}. $$

There are 3 real free parameter which can be interpreted as rotation angles and one imaginary phase $\delta$, which can cause CP-violation in the leptonic sector\footnote{In case of Majorana neutrinos two additional phases may arise.}.
The mixing matrix can be parameterized in the form

$$ \pmatrix{\nu_e \cr \nu_\mu \cr \nu_\tau } = \pmatrix{ 1 & 0 & 0 \cr 0 & c_{23} & s_{23} \cr
0 & -s_{23} & c_{23} \cr} \pmatrix{ c_{13} & 0 & s_{13}e^{i\delta} \cr 0 & 1 & 0 \cr -s_{13}e^{i\delta} & 0 & c_{13} \cr} 
\pmatrix{ c_{12} & s_{12} & 0 \cr -s_{12} & c_{12} & 0 \cr 0 & 0 & 1 \cr}
\pmatrix{\nu_1 \cr \nu_2 \cr \nu_3 \cr}. $$

Here, $s_{ij} = \sin \Theta_{ij}$ and $c_{ij} = \cos \Theta_{ij}$ with the rotation angles $\Theta_{ij}$.
 
An overview about the mechanism of neutrino oscillations can be found in many publications (e.g.~\cite{bilenky},\cite{lobo}).
The probability $P_e$ for electron-antineutrinos emitted from a reactor to survive in this state after a distance $L$ is
$P_e = 1 - \sin^2 2\Theta_{12} \sin^2 (1.267 \, {\Delta m^2_{12} \over eV^2} \, {L/m \over E/MeV}) $.   
Allowed values of the oscillation amplitude $\sin^2 2\Theta_{12}$ and the neutrino mass difference $\Delta m^2_{12}=m^2_2 - m^2_1$
are in agreement with the 'Large-Mixing-Angle'-(LMA) solution, which is the most probable parameter set when all solar neutrino experiments, including SNO, are analyzed. 
The best fit delivers the actual values~\cite{SNO-new}\\
$\Theta_{12} = 32.5^{+1.6}_{-1.7} \,\,\, deg $  and \\
$\Delta m^2_{12} = 7.1^{+1.0}_{-0.3} \times 10^{-5} \,\,\, eV^2$.\\
Full mixing (i.e. $\Theta_{12} = 45^o$) is excluded by 5.4$\sigma$.

In addition we have evidence for neutrino oscillations from atmospheric neutrino measurements performed in SuperKamiokande~\cite{atmo-SK} and other experiments (overview e.g.~\cite{atmo-noSK} and ref.\ therein).
Here the disappearance of muon-neutrinos $\nu_\mu$ is observed.
The best fit values here are $\Theta_{23} = 45 \,\, deg$ (i.e. full mixing) 
and $\Delta m^2_{23} = 2 \times 10^{-3} \,\, eV^2$.
Oscillation $\nu_\mu \to \nu_e$ with these parameters is excluded by the Chooz reactor neutrino experiment~\cite{Chooz}, which measured the flux at a distance of about 1km.
On the other hand the SuperKamiokande data indicate appearance of tau-neutrinos $\nu_\tau$ and disfavor the oscillation mode into sterile neutrinos~\cite{atmo-SK}.
In addition the disappearance of atmospheric $\nu_\mu$'s is going to be proven by the long-baseline K2K accelerator experiment\cite{k2k}.
Therefore it is widely believed, that $\nu_e \leftrightarrow \nu_\mu $-oscillations are observed in the solar neutrino experiments with parameters $\Theta_{12}$, $\Delta m^2_{12}$ as given above, and the anomaly in the atmospheric neutrino data is due to the oscillation mode $\nu_\mu \leftrightarrow \nu_\tau $ with the parameters $\Theta_{23}$, $\Delta m^2_{23}$.

The accelerator experiment LSND claimed evidence for neutrino oscillations~\cite{LSND} with values for the mass difference which are in the range 0.2 to 2eV$^2$.
A large, however not complete, parameter space was excluded by Karmen~\cite{Karmen} and former reactor experiments~\cite{Bugey}\cite{Goesgen}.
The LSND evidence is tested now in the recently started MiniBooNE experiment at Fermilab~\cite{MiniBooNE}.
If the LSND-evidence would be verified, a new 4th neutrino has to be introduced, which do not contribute to the $Z^0$-decay width, and is therefore considered to be sterile.
 
We conclude that from solar, reactor, and atmospheric neutrino measurements the coupling strengths of the mass differences between the first and the second family, as well as between the second and the third family have been determined.
However, we don't know yet the coupling parameter $\Theta_{13}$ and we don't know the absolute values of neutrino masses, as oscillation experiments only measure mass differences.

\section{Future solar neutrino experiments}

Despite the impressive success of SNO there is a vital interest in future solar Neutrino experiments aiming for neutrino spectroscopy at low energies.
Up to now the high abundant low energy part of the solar spectrum (i.e.\ the pp-, $^7$Be-, pep-, and CNO-neutrinos) has been measured only integral in the radiochemical Gallium experiments GALLEX/GNO~\cite{GNO} and SAGE~\cite{SAGE} and
direct experimental information on the strengths of the individual branches is still missing.

The transition of the rather high energy $^8$B-neutrinos is dominated by matter effects in the solar interior, whereas the low energy (E $<$ MeV) neutrinos undergo vacuum oscillations.
For those neutrinos the $\nu_e$-transition probability is expected to be only $\approx$1/3 
and not 2/3 as measured for $^8$B-neutrinos in SNO.   
The measurement of the transition strength in the low energy regime is therefore of great interest.

In addition the measurement of the dominating pp-, and $^7$Be-neutrinos would allow to determine important solar parameters with high precision in order to scrutinize stellar evolution theory.
In this context it is notable, that the actual results of solar $\nu$-experiments still would allow a CNO-contribution which exceeds the standard value by a factor $\approx 5$~\cite{bahcall}.
Future low energy experiments could improve this limit considerably.

Two funded experiments are aiming for low energy solar neutrino spectroscopy, especially for the first direct measurement of solar $^7$Be-neutrinos.
One is BOREXINO~\cite{BOREXINO} at the Italian Gran Sasso underground laboratory, the second is the already mentioned KamLAND experiment in the Japanese Kamioka mine.
Both would use neutrino electron scattering in a liquid scintillator as detection reaction.
Neutrinos arriving in the $\nu_\mu$-state would interact via this reaction with a probability of about 20\% compared to $\nu_e$'s.
Hence the observed interaction rate in the detectors (expected are $\approx$ 35 per day per 100t target mass) would tell how many $^7$Be-neutrinos changed their flavor due to oscillations when one compares this number with the calculated $^7$Be-flux from solar models.
The measurement is difficult, as the background in this low energy regime has to be suppressed substantially.
KamLAND has the advantage of a larger target mass (1000t compared to 300t in Borexino), hence more statistics and better self-shielding against external gamma and neutron background. 
On the other side BOREXINO has a lower cosmic muon flux due to the better shielding of the Gran Sasso laboratory.
In both experiments it has been shown, that the specifications on the intrinsic concentrations of U and Th in the liquid scintillator can be met~\cite{CTF}\cite{kl-lownu}.
However, the concentration of the long-lived Rn-daughter $^{210}$Pb and its radioactive daughters as well as the concentrations of radioactive Kr and Ar can be quite high, and special purification methods have been developed (see e.g.~\cite{low}\cite{ludwig}).
The achievable energy threshold will be probably in the range of about 250 keV due to the $^{14}$C background present in an organic scintillator.
KamLAND is already operating, measuring reactor neutrinos via the inverse beta decay on protons and
from the measurements of single events in the low energy regime it is clear, that the background below 1 MeV has to be lowered by $\approx$3 orders of magnitude~\cite{kl-lownu}.
The goal is to achieve this within the next two to three years.
Then, in additional tree to five years results from the solar neutrino phase of KamLAND could be expected.  
BOREXINO is still in the building-up period and delayed~\footnote{the delay is partly due to an accidental spill of about 50l of liquid scintillator in the environment which caused a formal stop of working with liquids. Work will be continued after further precautions are done.}
compared to the time plan, albeit most of the installations has been finished.
Successful measurements on low-level techniques with liquid scintillators have been performed in the Counting Test Facility of BOREXINO at Gran Sasso~\cite{CTF}\cite{pxe}.

Next generation solar experiments using noble gases (He, Ne, Xe) as neutrino target are under consideration.
In electron scattering experiments (XMASS~\cite{XMASS}, CLEAN~\cite{CLEAN}, HERON~\cite{HERON} and Super-MUNU~\cite{MUNU}) the threshold energy could be lower as in KamLAND and BOREXINO, as the $^{14}$C background is absent.
Hence even direct pp-$\nu$-detection could be feasible with these projects.
Charged current experiments in this field are considered in LENS~\cite{LENS} and MOON~\cite{MOON}.
The basic concept is the use of nuclei ($^{115}$In, $^{100}$Mo) as $\nu_e$ target with low energy thresholds
(e.g. $^{115}$In$ + \nu_e \to ^{115}$Sn$ + e^-$ with Q = 128 keV), where
the prompt electron allows neutrino spectroscopy.
With both isotopes delayed coincidence techniques can be applied, which helps to tag the neutrino event and to discriminate against background.

\section{Future reactor neutrino experiments}

The actual best fit values for oscillations with $\Delta m^2_{12} \approx 7 \times 10^{-5} \,\, eV^2$ would suggest to chose a distance for a future reactor experiment with a baseline shorter as it is realized in KamLAND.
With an average neutrino energy of about 4 MeV the optimal distance (i.e. half of the oscillation length)
would be $L_{osc}/2 = 1.24 \times {E/MeV \over \Delta m^2_{12}/eV^2} \approx 70 \,\, km$.
An experiment at this distance should observe the first minimum of the oscillation pattern in the energy spectrum with best accuracy and would improve the precision on the values of the oscillation parameters\cite{choubey}.

An reactor experiment at a distance of about 2 km would test on a possible disappearance with the atmospheric result of $\Delta m^2_{13} \approx 2 \times 10^{-3} \,\, eV^2$.
The actual best limit is coming from the Chooz experiment with $\sin^2 2\Theta_{13} \le 0.2$ at 90\%cl\cite{Chooz}.
An improvement of this constraint should be possible if two identical detectors are used.
The close one (distance $\approx 100m$) would monitor the neutrino flux and spectrum of the reactor with high statistics.
By comparing the data of both detectors most of the systematic uncertainties (e.g.\ neutrino cross section, neutrino flux, Uranium vs.\ Plutonium content of the fuel elements etc.) should cancel.
Possible sites with a large shielding (minimum $\approx 300$mwe) of at least the far detector are discussed in the USA, France, Russia, Brazil, and Japan.
The community is preparing a common paper on the subject, which will be published soon\cite{wp}.
Knowledge of $\Theta_{13}$ is important as a non-zero value is required for observing CP-violating and/or matter effects in future long baseline accelerator experiments.

\section{Direct neutrino mass searches}

The most sensitive direct search for neutrino masses is performed in Tritium endpoint measurements,
where a deviation in the electron spectrum close to the Q-value is searched for.
An upper limit on $m_\nu = \sum m_i \cdot \vert U_{ei} \vert ^2$ of 2.2eV has been achieved in the Mainz~\cite{mainz} as well as in the Troitsk experiment~\cite{troitsk} at 95\% cl.
This limit sets the scale on all neutrino masses $m_i$, as the oscillation results reveal that the mass differences are much smaller.
As the SNO-result can be explained via the matter enhanced oscillation effect it is evident that $m_2 > m_1$, as the matter effect only works if $m^2_{21}$ is positive.
However, the total neutrino mass hierarchy is not yet known.
In normal hierarchy the mass pattern would read $m_1 < m_2 < m_3$, whereas an inverted hierarchy would mean
$m_3 < m_1 < m_2$.
In addition it is unknown whether the masses are degenerated or not.
In the first case the mass splittings would be significantly lower as the absolute mass values.
With the future KATRIN project~\cite{katrin} at Karlsruhe, Germany, a sensitivity down to $\sim 0.35$ eV is aimed.

Limits on the sum of neutrino masses come from astrophysical and cosmological observations.
By employing the Sloan Digital Sky Survey, the Two Degree Field Galaxy Redshift Survey and the cosmic microwave background data from the Wilkinson Microwave Anisotropy Probe a limit of 0.75eV (at two sigma) and 1.1 eV at three sigma has been reported recently~\cite{astro}.  

The search for the neutrinoless double beta ($\beta\beta 0\nu$) decay allows to test the nature of the neutrino and to search for neutrino masses.
The decay is searched for in rare weak processes of the type $A(Z) \to A(Z+2) + 2e^-$, where the single beta-decay of the nucleus $A(Z)$ is prohibited kinematically.
In this process the neutrino is a virtual particle emitted as a right-handed anti-particle in one vertex and absorbed as left-handed particle in the second vertex.
Hence, condition for the ($\beta\beta 0\nu$)-decay to happen is a Majorana neutrino (particle = anti-particle) which implies Lepton number violation $\Delta L = 2$ and non-zero neutrino masses.
The ($\beta\beta 0\nu$)-decay is sensitive to the so-called effective neutrino mass
$m_{ee} = \vert \sum U^2_{ei} \cdot m_i \vert$, which is a coherent sum over all mass eigenstates.
The values $U^2_{ei}$ may comprise complex phases, which could lead to partial cancelation of different terms in the sum.
Uncertainties in the nuclear matrix elements of ($\beta\beta 0\nu$)-decays still contribute to the uncertainty in $m_{ee}$ by about a factor of 2.
The best limits on $m_{ee}$ are coming from experiments using large masses of enriched $^{76}$Ge-detectors.
A signal for the ($\beta\beta 0\nu$)-decay would show up as a line in the energy spectrum at the endpoint of 2039 keV.
The Heidelberg-Moscow experiment at the Gran-Sasso underground laboratory published a limit of $m_{ee} < 0.35$eV at 90\%cl~\cite{hm}.
Due to uncertainties in the matrix element this limit is interpreted to be in the range of about 0.3 to 1.0 eV~\cite{cremo}.
The IGEX experiment reports about a limit of $m_{ee} < $0.33 - 1.35 eV~\cite{igex}.

In 2002 few members of the Heidelberg-Moscow collaboration claimed evidence for ($\beta\beta 0\nu$)-decay with a best fit value for the half-lifetime $T_{1/2} = 1.5 \cdot 10^{25}$y~\cite{klapdor}, corresponding to $m_{ee} = 0.11 - 0.56$eV.
The result is based on a renewed analysis of the data using a peak detection mode by narrowing the fit interval around the region of interest.
The paper has been criticized as the authors could not assign neighbored lines to known background contributions.
In future one may hope, that high sensitive experiments will test this claim of evidence.
If verified, it would be a further extremely important result which was obtained in low energy neutrino physics.
It would not only settle the absolute neutrino mass scale~\footnote{which would show a degenerate behavior}, but would also reveal the nature of the neutrino to be a Majorana particle.
There exist a large number of proposals for future double beta experiments for several nuclei as candidates.
An overview about prospects in this field is given e.g.\ in~\cite{cremo}.

\section{Low Energy Neutrino Astronomy}

With the observation of solar neutrinos, pioneered by R.\ Davis with the Homestake experiment, and the observation of neutrinos emitted in a Supernova in February 1987 a new window was opened in Astronomy.
Neutrinos can be used as probes in order to receive information from astrophysical objects, which otherwise would be unaccessible.
This line is followed by the development of large neutrino telescopes for very high energies which are going to be build in water or ice as Cherenkov detectors.

In this paper we propose a large liquid scintillator detector for low energy neutrino astronomy (LENA).
With LENA one may aim at important topics like
time resolved flavor specific detection of galactic supernova neutrinos, supernova relic neutrinos, high statistic solar neutrino spectroscopy, detection of terrestrial neutrinos, long baseline neutrino experiments, and proton decay.
With this scientific program fundamental aspects in particle astrophysics as well as elementary particle and geophysics would be addressed. 
Here we present the detector characteristics of LENA and discuss its potential for supernova physics, solar physics, and proton decay searches.    

\subsection{Detector Characteristics}

The detector is proposed to consist of a large volume liquid scintillator with cylindrical shape, approximately 30m in diameter and 90m in length, equipped with a photomultiplier (PM) coverage of about 30\%. This can be achieved with about 12000 PMs with 50cm diameter each. The detector could be placed for instance under sea close to the coast at Pylos (Greece) at the deepest site in Europe ($\approx$5000m) or at the center of underground physics in Pyh\"asalmi (CUPP, Finland) in 1400 m depth of rock ($\approx$4060mwe). Both sites are favored as being far away from nuclear power plants, which may significantly contribute to the $\bar{\nu_e}$ background in the search for relic supernovae neutrinos. 

We propose PXE (phenyl-o-xylylethane) as scintillator for LENA. It has been investigated in the R\&D of BOREXINO\cite{pxe}. PXE has a high density of 0.99g/cm$^3$ and shows a high light yield of about 88\% relative to pseudocumene. According to UN regulations PXE is legally non-hazardous for transportation purposes. It is safe to handle due to its high flash-point of $145^o$C. A light attenuation length of about 12m at 450nm wavelength has been achieved\cite{stefan}. 
We estimate a photoelectron yield of about 120 pe/MeV for a beta-like event which occurs in the center of LENA. Hence, for the discussion below an energy resolution $\delta E / E = 0.1 (E/MeV)^{-1/2}$ can be assumed. A position resolution $\delta r = 25cm (E/MeV)^{-1/2}$ for single events can be expected. After purification in Si-gel columns the mass-concentrations of
$^{238}$U and $^{232}$Th has been measured via NAA to be below $10^{-17}$ and $2 \times 10^{-16}$, respectively\cite{roger}. 
We conclude, that with PXE a non-hazardous, pure scintillator with high light yield and large attenuation length would be available for LENA.

\subsection{Supernova Neutrinos}
\subsubsection{Galactic Supernova Neutrino Detection}

In case of a supernova at the center of our galaxy $\approx$15000 $\nu$-events in LENA can be expected. Using an organic scintillator containing $^{12}$C allows the distinct flavor specific neutrino and antineutrino detection by the following reactions:\\ 
$1)\,\, \bar{\nu_e} + p \to e^+ + n$ (Q = 1.8 MeV),\\
$2)\,\, \bar{\nu_e} + ^{12}C \to e^+ + ^{12}B$ (Q = 17.3 MeV),\\
$3)\,\, \nu_e + ^{12}C \to ^{12}N + e^- $ (Q = 13.4 MeV),\\
$4)\,\, \nu_x + ^{12}C \to ^{12}C^* + \nu_x$ with $ ^{12}C^* \to ^{12}C + \gamma$ (E$_\gamma$ = 15.1 MeV) and\\
$5)\,\, \nu_x + p \to \nu_x + p$ (elastic scattering).\\
The spectral $\bar{\nu_e}$-contribution can be identified via the first cc-reaction, utilizing the delayed coincidence between the prompt positron and the succeeding neutron capture on hydrogen. This is the dominant reaction mode with the highest cross section and one would expect ca.\ 7000 events. 

With this information the $\nu_e$-spectrum above 13.4 MeV can be disentangled from the cc-reactions (2) and (3) which yield an event number of $\approx$500 and $\approx$100, respectively. 
Both reactions can be tagged by the re-decay of the daughter nuclei 
$^{12}$B ($\beta^-$, $T_{1/2} = 20 ms$) and $^{12}$N ($\beta^+$, $T_{1/2} = 11 ms$). 
All $\nu$-flavors participate in the nc-interaction (4), which yields information about the total SN-$\nu$ flux.
Here about 4000 events may be expected. 
Elastic scattering of all flavors on hydrogen (5) will lead to an intense ($\approx$2200 events) low energy signal due to recoil protons\cite{beacom}. Observation of this interaction type becomes feasible as the detector threshold can be as low as $\approx$0.2 MeV (equivalent to ca.\ 25 pe for an event in the center).
This signal can be clearly separated from the 'high energy' pulses from reactions (1) to (4). The measured proton recoil spectrum reflects the incoming SN-$\nu$ spectrum. If the mean energies of $\nu_\mu$, $\nu_\tau$ are above the mean energy of $\nu_e$, the signal above threshold should be dominated by muon- and tau-$\nu$`s. 
In addition, elastic $\nu$-scattering off electrons would be observed. This signal would yield a low-energy supplement of the events seen in large water Cherenkov detectors like SuperKamiokande.

Observation of such a burst would allow to measure the time development of the specific $\nu$-fluxes in a Supernova and would reveal important details of the explosion mechanism.
In addition the $\bar{\nu_e}$ energy spectrum would show wiggles which are caused by oscillation matter effects when the neutrinos cross the Earth before entering the detector.
These wiggles are observable in LENA due to the good energy resolution and statistics, and would reveal information about $\nu$-oscillation parameters as well as the mass hierarchy (for details see \cite{georg}). 

\subsubsection{Supernova Relic Neutrino Detection}

It is generally believed that supernova core-collapses have traced the star formation history in the Universe. In these explosions a great number of SRN-$\nu$`s must have been emitted. The comparison of the experimentally observed SRN-spectrum with the predicted results of models will deliver valuable information on the star formation history in the Universe. Current models on the star formation rate contain various uncertainties, especially at high redshift regions (see e.g.\cite{ando} and refs.\ therein). The supernova rate is expected to be proportional to the star formation rate as the lifetime of progenitors of core-collapse supernovae is much shorter than the cosmological time scale. 

The SRN-flux determination is one of the targets of LENA. 
The currently best limit on SRN comes from the SuperKamiokande detector\cite{snr} giving an upper limit of 
1.2 cm$^{-2}$s$^{-1}$ for $\bar{\nu_e}$ with a threshold of 19.3 MeV. 
With LENA the sensitivity for the SRN search should be drastically improved as the delayed coincidence between the prompt positron and the captured neutron in the inverse beta reaction can be utilized. 
This strongly reduces the background and it should be possible to reach an energy threshold of $\approx$9 MeV. 
A lower threshold is prohibited by the ubiquitous $\bar{\nu_e}$`s from nuclear power plants.
According to the most recent models\cite{ando} the predicted SRN flux at 10 MeV is about 0.6 cm$^{-2}$sec$^{-1}$MeV$^{-1}$ and LENA should observe an event rate of about 4/year.

\subsection{Solar Neutrinos}

One expects, that BOREXINO and KamLAND will measure the solar $^7$Be-$\nu$ flux. 
However, the fluxes of pep- as well as CNO-$\nu$ are faint and the unavoidable background due to cosmic rays will make the measurement very difficult (details in \cite{tanja}).
With LENA the solar $\nu$-rates would be $\sim 5400$/d in a fiducial volume of about 22000m$^3$ for 
$^7$Be-$\nu$, 150/d for pep-$\nu$, and 210/d for CNO-$\nu$. The high statistics would help in the $\nu$-signal identification as the annual 7\% change due to the eccentricity of the Earth's orbit should be observable.

The $^7$Be-$\nu$ rate could be determined with an accuracy of about 5\% after only one year of measurement. 
From this measurement together with the knowledge of the solar luminosity and $\nu$-oscillation parameters the fundamental pp-flux can be determined with an accuracy of better than 0.5\%\cite{pena}. 
Important astrophysical parameters like the R$_{34}$/R$_{33}$ branching ratio could be measured to a precision of about 1\%\cite{pena}. 

Due to the MSW-effect the survival probability of $\nu_e$`s created in the solar center depends on the energy and on the density profile of the sun. The high-statistic measurement of $^7$Be-$\nu$`s with LENA would allow to test on temporal fluctuations of the solar density profile with high precision. Such temporal density fluctuations could be created by solar g-mode waves, which are not observed so far by helioseismology. Following the arguments from \cite{bala} a density fluctuation of 1.5\% should result in a $^7$Be-$\nu$ flux change of about 10\%.

\subsection{Geoneutrinos}

The thermal heat flow emitted by the Earth is small and is measured to by about 80 mW/m$^2$.
However, the integral terrestrial heat is around 40 TW (uncertainty $\sim$20\%), which corresponds to the power of about 10,000 nuclear reactor plants.
What is the source of this energy?
How much is the contribution of radioactivity in the Earth?
Those questions are not understood quantitatively, but by the 
measurements of geo-neutrinos, which are emitted in beta decays in the Uranium and Thorium chains, the determination of the radiogenic contribution should be possible~\cite{krauss}.
Those geo-neutrinos with energies above 1.8 MeV could be detected by the inverse beta decay $\bar{\nu_e} + p \to e^+ + n$.
The spectrum of geo-neutrinos is below $\sim$3 MeV and hence one can distinguish them from reactor neutrinos which have energies up to about 8 MeV.
Indeed the KamLAND collaboration used for their analysis of oscillations of reactor neutrinos only the spectrum above 2.6 MeV, as the contribution of geo-neutrinos below this energy to the total spectrum is unknown to a large extent.
In future publications, when statistics is improved, one may expect that KamLAND achieves a first measurement of geo-neutrinos by extrapolating the measured reactor spectrum to the low energy region.
The flux of geo-neutrinos should depend significantly on the site of the detector.
Therefore it would be advantageous, if more detectors at different places would work.
With a large detector like LENA one would be able to measure also the flux of geo-neutrinos coming from the mantle of the Earth and hence the abundances of Uranium and Thorium in the mantle could be calculated.
In LENA the event rate of geo-neutrinos are estimated to be in the range of (600 - 3000)/year.
The background due to reactor neutrinos can be determined as described above and is small for the aimed detector sites of LENA.
After a measuring period of 3 years an accuracy of about 3\% can be expected. 

\subsection{Atmospheric Neutrinos}

With LENA the low energy part of atmospheric neutrinos could be explored in more detail.
The quasi-elastic reactions $\bar{\nu_e} + p \to e^+ + n$ and $\bar{\nu_\mu} + p \to \mu^+ + n$ could be used to measure their flux in the energy region between $\sim$100 MeV and 1 GeV.
As the $\mu^+$ decays after 2.2 $\mu s$ the $\bar{\nu_\mu}$ would be tagged by a threefold delayed coincidence and one could separate those events clearly from $\bar{\nu_e}$-events.
Without oscillations the ratio between both event rates should be $\bar{\nu_e} / \bar{\nu_\mu} = 0.5$. 

At 200 MeV the oscillation length of $\bar{\nu_e}$ due to $\Delta m^2_{12}$ (the 'solar mass splitting') should be around 7000 km and this should lead to oscillations of the type $\bar{\nu_e} \leftrightarrow \bar{\nu_\mu}$.
At 1 GeV however the oscillation length is at 35000 km and the oscillation probability should become significantly smaller.
On the other hand $\bar{\nu_\mu}$-oscillation into $\bar{\nu_\tau}$ due to the $\Delta m^2_{23}$ splitting should be always present and lead to a depletion of the $\bar{\nu_\mu}$-flux.
This implies, that an well defined energy dependent ratio $\bar{\nu_e} / \bar{\nu_\mu}$ due to the various oscillations should be observable in LENA. 

\subsection{Long baseline experiment with LENA}

In case a high energy neutrino beam is directed to LENA the detector could be used also for the long-baseline oscillation studies including the matter effects and the search for CP-violation in the leptonic sector.
Muon events should be separable from electron events by their different lengths in the detector.
For this purpose the axis of the detector should be parallel to the neutrino beam.
In addition the decay $\mu^+ \to e^+ \bar{\nu_\mu} \nu_e$ can be used to tag muon events.
It is interesting to note, that LENA at Pylos would be off-axis in the elongated direction of the aimed CERN to Gran Sasso beam.
Further studies of the detector capabilities for this high-energy studies are en route. 

\subsection{Proton Decay}

In the search for the decay mode $p\to K^+ \nu$ which is favored by SUSY models, water Cherenkov detectors are limited as the energy of the Kaon is below the Cherenkov threshold.
In  LENA this decay mode is visible. 
With a probability of 63.5\% the Kaon decays via $K^+ \to \mu^+ \nu_\mu$ and in this case the scenario for a signal in LENA would be:\\
i) a prompt mono-energetic K$^+$ (T=105 MeV),\\
ii) a short delayed ($\tau$=12.8 ns) mono-energetic $\mu^+$ (T=152 MeV),\\
iii) a long delayed ($\tau$=2.2 ms) $e^+$ from the following $\mu^+$ decay.

With a probability of 21.2\% the Kaon decays via $K^+ \to \pi^+ \pi^0$ and in this case the scenario for a signal in LENA would be:\\
i) a prompt mono-energetic K$^+$ (T=105 MeV),\\
ii) a short delayed mono-energetic $\pi^+$ (T=108 MeV) accompanied by an electromagnetic shower due to the 2-$\gamma$ decay of the $\pi^0$ (E=246 MeV),\\
iii) a short delayed ($\tau$=26 ns) mono-energetic $\mu^+$ with T = 4 MeV from the $\pi^+$ decay,\\
iv) a long delayed ($\tau$=2.2 ms) $e^+$ from the $\mu^+$ decay.

Due to the good energy resolution, the fast detector response, and position reconstruction the search for the proton decay into this channel should be performed basically background free with a high efficiency.
For LENA the reachable sensitivity for the proton decay  $p\to K^+ \nu$ could be close to a lifetime limit between 
$10^{34}$ and $10^{35}$ years after a measuring time of 10 years. The minimal SUSY SU(5) model predicts the decay mode     to be dominant with a partial lifetime varying from $10^{29}$ to $10^{35}$ years\cite{pdectheo}.
The actual best limit on this decay mode from SuperKamiokande is $6.7 \times 10^{32}$ y (90\% cl)\cite{pdec}.
LENA could detect further interesting nucleon decay modes, which are 'invisible' for Cherenkov detectors.
More details can be found in~\cite{swoboda}.

\end{document}